\documentclass[superscriptaddress,twocolumn,nofootinbib]{revtex4}
\usepackage{amssymb}
\usepackage{amsfonts}
\usepackage{amsmath}
\usepackage{graphicx}
\usepackage{dcolumn}
\usepackage{bm}
\usepackage[latin1]{inputenc}
\usepackage[dvips]{hyperref}

\begin{document}

\title{Remnant group of local Lorentz transformations in $f(T)$ theories}
\author{Rafael Ferraro}
\email[Member of Carrera del Investigador Cient\'{\i}fico (CONICET,
Argentina); ]{ferraro@iafe.uba.ar} \affiliation{Instituto de
Astronom\'ia y F\'isica del Espacio (IAFE, CONICET-UBA), Casilla de
Correo 67, Sucursal 28, 1428 Buenos Aires, Argentina.}
\affiliation{Departamento de F\'isica, Facultad de Ciencias Exactas
y Naturales, Universidad de Buenos Aires, Argentina.}
\author{Franco Fiorini}
\email[Member of Carrera del Investigador Cient\'{\i}fico (CONICET,
Argentina); ]{francof@cab.cnea.gov.ar} \affiliation{Instituto
Balseiro, Universidad Nacional de Cuyo, R8402AGP Bariloche,
Argentina.} 

\begin{abstract}
It is shown that the extended teleparallel gravitational theories,
known as $f(T)$ theories, inherit some \emph{on shell} local Lorentz
invariance associated with the tetrad field defining the spacetime
structure. We discuss some enlightening examples, such as Minkowski
spacetime and cosmological (Friedmann-Robertson-Walker and Bianchi type I) manifolds. In
the first case, we show that the absence of gravity reveals itself
as an incapability in the selection of a preferred parallelization
at a local level, due to the fact that the infinitesimal local
Lorentz subgroup acts as a symmetry group of the frame
characterizing Minkowski spacetime. Finite transformations are also
discussed in these examples and, contrary to the common lore on the
subject, we conclude that the set of tetrads responsible for the
parallelization of these manifolds is quite vast and that the
remnant group of local Lorentz transformations includes one and two
dimensional Abelian subgroups of the Lorentz group.
\end{abstract}

\maketitle

\section{Introduction}\label{sec:intro}

In spaces with absolute parallelism, the geometry of a given spacetime is
encoded in the tetrad field $\mathbf{e}^{a}$. This global basis of the
tangent bundle constitutes a preferred reference frame which defines the
spacetime structure $(\mathcal{T}(\mathcal{M}),\mathbf{e}^{a})$. In general,
any tetrad $\mathbf{e}^{a}$ which serves as a global frame leads to a
certain Lorentzian geometry, characterized by $\mathbf{g}=\eta _{ab}~\mathbf{%
e}^{a}\otimes \mathbf{e}^{a}$. The common belief concerning the
geometrical structure of gravitational theories in such spaces, is
based on the notion of absolute parallelism characterizing
Weitzenb\"{o}ck spacetime. According to this description, a preferred
reference frame emerges as the agent which defines the spacetime
structure by means of a parallelization process. In principle, the
notion of parallelism so obtained should be defined only with
arbitrariness of making global Lorentz transformations of the
preferred frame, for this special tetrad field dictates what an
autoparallel is: a curve will be autoparallel if its tangent vector
has constant components with respect to the global preferred frame.
However, from a purely mathematical point of view, it has been known for a long time that if a given space is parallelizable, the
vector fields carrying out such a parallelization are not unique \cite{Steenrod}. On physical grounds, and for some gravitational
theories constructed out of the concept of absolute parallelism (like $f(T)$ gravity, the one concerned in the present work), this means that, apart from the freedom to perform \emph{global}
Lorentz transformations to a given global frame, certain
\emph{local} boosts and rotations will act as a symmetry group of
the theory.

Among the gravitational theories relaying on absolute parallelism,
the so called $f(T)$ gravity
\cite{Nos1,Nos2,Bengochea:2008gz,Linder:2010py} has been the object
of considerable study in the last few years (see, for instance
\cite{Nosal,Izumi,Geng,Boehmer} and references contained therein).
Since the very beginning, it was realized that the local Lorentz
symmetry is not present in these theories \cite{Nos1},
\cite{Barrow1,Barrow2,Barrow3}, and as a consequence of this,
preferred reference frames emerge as the agent encoding the
gravitational field \cite{Nos3,Nos4}. It is our concern now to show
that, besides the global symmetry always present in any theory
constructed upon the notion of absolute parallelism, these preferred
frames are defined with the arbitrariness of making certain \emph{local}
Lorentz transformations. The admissible group of local Lorentz
transformations depends on the particular spacetime under
consideration: for a given frame $\mathbf{e}^a$ representing a
solution of the $f(T)$ motion equations, there exist a subgroup
$\mathcal{A}(\mathbf{e}^a)$ of the Lorentz group which officiates as
a symmetry group. The presence of a restricted local invariance of
this sort have been occasionally documented in the literature, see for instance ref. \cite{Hayashi82} regarding the theory exposed in
\cite{Hayashi79}.

The study of the group $\mathcal{A}(\mathbf{e}^a)$ is mandatory for
at least two important reasons. On one hand, the knowledge of
$\mathcal{A}(\mathbf{e}^a)$ allows us to obtain new solutions of the
motion equations from the old. This is particularly important if they
involve the matching of different tetrads, as it happens in stellar
and wormhole models, where we have two different spacetimes which
must be smoothly matched on a certain hypersurface. On the other
hand, in order to perform a correct counting of degrees of freedom, detailed information about the symmetries of the theory under
consideration becomes fundamental. These two constitute the main
motivations of the present work, and it is expected that the
techniques involved in this article might serve for answering
similar questions in other theories relying on absolute
parallelism, for instance, in Born-Infeld gravity \cite{Nos5,Nos6},
and in the extensions of Gauss-Bonnet gravity in the teleparallel
context \cite{Sar1,Sar2}.

In order to understand the nature of $\mathcal{A}(\mathbf{e}^a)$ in
the context of $f(T)$ gravity, we first set down the preliminary
geometrical concepts in section \ref{setting}. After these
ingredients are presented there, we expose General Relativity (GR)
and its Teleparallel Equivalent (TEGR) in section \ref{EHTEGR}. The
behavior of $f(T)$ theories under local Lorentz transformation is
then thoroughly discussed in section \ref{fdet}, followed by a
number of important examples which crystallize the concepts of the
former sections. These examples are the central point of section
\ref{sec:infini}. Finally, we establish our conclusions in
\ref{sec:conclusion}.

\section{Geometrical Setting}\label{setting}

The theories where gravity is regarded as the geometry of the spacetime rest
on two basic concepts of differential geometry: \textit{torsion} $\mathbf{T}%
^{i}$ and \textit{curvature} $\mathbf{R}_{\ j~}^{i}$,
\begin{equation}
\mathbf{T}^{i}\ \doteq \ D\mathbf{E}^{i}\ \doteq \ d\mathbf{E}^{i}+\mathbf{%
\omega }_{\ j}^{i}\wedge \mathbf{E}^{j}\ .  \label{torsion}
\end{equation}%
\begin{equation}
\mathbf{R}_{\ j}^{i}\ \doteq \ d\mathbf{\omega }_{\ j}^{i}+\mathbf{\omega }%
_{\ k}^{i}\wedge \mathbf{\omega }_{\ j}^{k}\ .  \label{curvature}
\end{equation}%
Torsion $\mathbf{T}^{i}$ is the covariant derivative of the 1-forms
constituting a local basis $\{\mathbf{E}^{i}\}$ of the cotangent space. The
covariant derivative $D$ is defined by endowing the manifold with a \textit{%
spin connection}, which is a set of 1-forms $\mathbf{\omega }_{\ j}^{i}$
taking care of additional tensor characteristics of the object under
differentiation. $D$ is an exterior derivative on $p-$forms preserving their
tensor-valued features. For instance, $\mathbf{T}^{i}$ is a vector-valued
2-form; it transforms as $\mathbf{T}^{i^{\prime }}=\Lambda _{\ \
i}^{i^{\prime }}\,\mathbf{T}^{i}$ under the change of basis $\mathbf{E}%
^{i^{\prime }}=\Lambda _{\ \ i}^{i^{\prime }}\,\mathbf{E}^{i}$. This is so
because the spin connection transforms as
\begin{equation}
\mathbf{\omega }_{\ j^{\prime }}^{i^{\prime }}\ =\ \Lambda _{\ i}^{i^{\prime
}}\ \mathbf{\omega }_{\ j}^{i}\ \Lambda _{\ j^{\prime }}^{j}+\Lambda _{\
k}^{i^{\prime }}\ d\Lambda _{\ j^{\prime }}^{k}\   \label{transformation}
\end{equation}%
(matrices $\Lambda _{\ i^{\prime }}^{i}$, $\Lambda _{\ i}^{i^{\prime }}$ are
inverses of each other; the dual basis in the tangent space transforms as $%
\mathbf{E}_{i^{\prime }}=\Lambda _{\ i^{\prime }}^{i}\,\mathbf{E}_{i}$).%
Analogously, $\mathbf{R}_{\ j}^{i}$ is tensorial in the indices $i$,
$j$. However $\mathbf{R}_{\ j}^{i}$ cannot be thought of as the
covariant derivative
of $\mathbf{\omega }_{\ j}^{i}$ because the connection is not a tensor. $%
\mathbf{R}_{\ j}^{i}$ can be covariantly differentiated to obtain the
(second) Bianchi identity,
\begin{equation}
D\mathbf{R}_{\ j}^{i}\ =\ d\mathbf{R}_{\ j}^{i}+\mathbf{\omega }_{\
k}^{i}\wedge \mathbf{R}_{\ j}^{k}-\mathbf{\omega }_{\ j}^{k}\wedge \mathbf{R}%
_{\ k}^{i}\ \equiv \ 0\ .
\end{equation}%
Besides, by differentiating the torsion we obtain the first Bianchi identity:%
\begin{equation}
D\mathbf{T}^{i}-\mathbf{R}_{\ j}^{i}\wedge \mathbf{E}^{j}\equiv \ 0\ .
\end{equation}%
In gravitational theories of geometrical character, we choose an
orthonormal basis or tetrad $\{\mathbf{e}^{a}=e_{\mu }^{a}\ dx^{\mu
}\}$ and the spin connection $\{\mathbf{\omega }_{\ \ b}^{a}\} $ to
play the role of potentials for describing the gravitational fields
(torsion and curvature). The assumed orthonormality of the tetrad
establishes the link tetrad-metric:
\begin{equation}
\eta ^{ab}=g^{\mu \nu }\ e_{\mu }^{a}\ e_{\nu }^{b}\ ,\hspace{0.3in}\mathbf{g%
}=\eta _{ab}\ \mathbf{e}^{a}\otimes \mathbf{e}^{b}.  \label{metric}
\end{equation}%
This link is invariant under local Lorentz transformations $\mathbf{e}%
^{a^{\prime }}=\Lambda _{\ \ b}^{a^{\prime }}(x)\,\mathbf{e}^{b}$ (i.e.,
those linear transformations preserving orthonormality). On the other hand
the spin connection is assumed to be \textit{metric}, which means the
vanishing of the covariant derivative of the \textit{Lorentz} tensor-valued
0-form $\eta _{ab}$:
\begin{equation}
0\ =\ D\eta _{ab}\ =\ d\eta _{ab}-\mathbf{\omega }_{\ \ a}^{c}\ \eta _{cb}-%
\mathbf{\omega }_{\ \ b}^{c}\ \eta _{ac}\ ,
\end{equation}%
i.e.,
\begin{equation}
\mathbf{\omega }_{ba}\ =\ -\mathbf{\omega }_{ab}  \label{metricity}
\end{equation}%
(Lorentz tensor indexes $a,b,...$ are lowered with $\eta _{ab}$). This
property also implies
\begin{equation}
D\epsilon _{abcd}\ =\ 0\ ,  \label{metricity2}
\end{equation}%
where $\epsilon _{abcd}$ is the Levi-Civita symbol, which is a tensor under
Lorentz transformations.

General Relativity is a theory of gravity where the connection is metric and
torsionless; it is the Levi-Civita connection $\overset{L}{\mathbf{\omega }}%
{}_{\ j}^{\,i}$:
\begin{equation}
d\mathbf{E}^{i}+\overset{L}{\mathbf{\omega }}{}_{\ j}^{\,i}\wedge \mathbf{E}%
^{j}=0\ ,\hspace{0.3in}\overset{L}{\mathbf{\omega }}_{ba}=-\overset{L}{%
\mathbf{\omega }}_{ab}\ .  \label{Levi-Civita}
\end{equation}%
These relations can be solved for the Levi-Civita connection in terms of the
exterior derivative of the tetrad:
\begin{equation}
\left( \overset{L}{\mathbf{\omega }}_{ab}\right) _{c}=\frac{1}{2}\ \left[
\left( d\mathbf{e}_{a}\right) _{bc}+\left( d\mathbf{e}_{b}\right)
_{ca}-\left( d\mathbf{e}_{c}\right) _{ab}\right] \ .  \label{LC-components}
\end{equation}%
For connections differing from the Levi-Civita connection it is convenient
to introduce the contorsion as the set of 1-forms expressing such a
difference:
\begin{equation}
\mathbf{K}_{\;j}^{i}\doteq \mathbf{\omega }_{\ j}^{i}-\overset{L}{\mathbf{%
\omega }}{}_{\ j}^{\,i}\ .  \label{contorsion}
\end{equation}%
Although connections are not tensors, the nontensorial term in the
transformation (\ref{transformation}) is equal for any connection. Therefore
the difference between connections is a tensor. Some useful properties of the
contorsion tensor can be consulted in the appendix \ref{propcontor}.

\section{Einstein-Hilbert and TEGR Lagrangians}\label{EHTEGR}

In this Section we will suppress the symbol of wedge product, since no
confusion exists provided that the order between $p-$forms is preserved.

Einstein-Hilbert Lagrangian is the Lorentz scalar-valued 4-form defined as
\begin{equation}
L_{_{EH}}\ =\ \frac{1}{4\,\kappa }\ \epsilon _{abcd}\ \mathbf{e}^{a}\
\mathbf{e}^{b}\ \overset{L}{\mathbf{R}}{}^{cd}\ ,  \label{EH}
\end{equation}%
where $\kappa =8\,\pi \,G$. Property 4 of appendix \ref{propcontor} implies that%
\begin{widetext}
\begin{eqnarray}
L_{_{EH}}\ &=&\ \frac{1}{4\,\kappa}\ \epsilon _{abcd}\ \mathbf{e}%
^{a}\, \mathbf{e}^{b}\, \left( \mathbf{R}^{cd}-\overset{L}{D}\mathbf{%
K}^{cd}-\mathbf{K}_{\ e}^{c}\, \mathbf{K}^{ed}\right)  \label{EH2} \\
\ &=&\ \frac{1}{4\,\kappa}\ \left[ \epsilon _{abcd}\
\mathbf{e}^{a}\, \mathbf{e}^{b}\, \left(
\mathbf{R}^{cd}-\mathbf{K}_{\ e}^{c}\,
\mathbf{K}^{ed}\right) -\overset{L}{D}\left( \epsilon _{abcd}\ \mathbf{e}%
^{a}\, \mathbf{e}^{b}\, \mathbf{K}^{cd}\right) \right] ,  \notag
\end{eqnarray}%
\end{widetext}where we have used that Levi-Civita connection is metric ($%
\overset{L}{D}\!\,\epsilon _{abcd}=0$) and torsionless ($\overset{L}{D}\!\,%
\mathbf{e}^{a}=0$). Moreover, $\overset{L}{D}\left( \epsilon _{abcd}\
\mathbf{e}^{a}\ \mathbf{e}^{b}\ \mathbf{K}^{cd}\right) =d\left( \epsilon
_{abcd}\ \mathbf{e}^{a}\ \mathbf{e}^{b}\ \mathbf{K}^{cd}\right) $, because $%
\epsilon _{abcd}\ \mathbf{e}^{a}\ \mathbf{e}^{b}\ \mathbf{K}^{cd}$ is a
Lorentz scalar. So the last term in (\ref{EH2}) is a boundary term that can
be suppressed:
\begin{equation}
L\ =\ \frac{1}{4\,\kappa }\ \epsilon _{abcd}\ \mathbf{e}^{a}\ \mathbf{e}%
^{b}\ \left( \mathbf{R}^{cd}-\mathbf{K}_{\ e}^{c}\ \mathbf{K}^{ed}\right) .
\label{EH3}
\end{equation}%
The Lagrangian (\ref{EH3}) now contains an arbitrary connection $\mathbf{%
\omega }^{cd}$; however, it does not provide any dynamics for $\mathbf{\omega
}^{cd}$. In fact, the Lagrangian (\ref{EH3}) is the Einstein-Hilbert
Lagrangian (\ref{EH}) modulo a boundary term. Since $\mathbf{\omega }^{cd}$
is not contained in the Einstein-Hilbert Lagrangian, we conclude that the
variation of (\ref{EH3}) with respect to $\mathbf{\omega }^{cd}$ will
produce a boundary term to compensate for the variation of the suppressed
boundary term appearing in (\ref{EH2}):
\begin{equation}
\delta _{\mathbf{\omega }}L\ \ =\ \frac{1}{4\,\kappa }\ d\left( \epsilon
_{abcd}\ \mathbf{e}^{a}\ \mathbf{e}^{b}\ \delta \mathbf{\omega }^{cd}\right)
.
\end{equation}%
So $\mathbf{\omega }^{cd}$ enters the Lagrangian (\ref{EH3}) as a \textit{%
dummy} variable to be chosen in an arbitrary way. TEGR chooses $\mathbf{%
\omega }^{cd}$ to be zero, which is the Weitzenb\"{o}ck connection (for the form
Weitzenb\"{o}ck connection acquires in a coordinate basis, see the Appendix \ref{slang}).
So, $\mathbf{R}^{cd}$ vanishes and $\mathbf{K}_{\ e}^{c}=-\overset{L}{%
\mathbf{\omega }}{}_{\ e}^{c}[\mathbf{e]}$ becomes linear and homogeneous in
derivatives of the tetrad. Then
\begin{equation}
L_{_{TEGR}}\ =\ -\frac{1}{4\,\kappa }\ \epsilon _{abcd}\ \mathbf{e}^{a}\
\mathbf{e}^{b}\ \mathbf{K}_{\ e}^{c}[\mathbf{e]\ K}^{ed}[\mathbf{e]}\ .
\label{LTEGR}
\end{equation}%
Thus, the freezing of $\mathbf{\omega }^{cd}$ throws the Lagrangian into a
form quadratic in first derivatives of the tetrad (see (\ref{LC-components}%
)). However, we cannot freeze a connection without paying a price. Although $%
\mathbf{\omega }^{cd}$ is a dummy dynamical variable in (\ref{EH3}), it
plays the important role of making (\ref{EH3}) a Lorentz scalar-valued
volume (i.e., (\ref{EH3}) is invariant under local Lorentz transformations
of the tetrad). This is because $\mathbf{K}_{\ e}^{c}$ is a Lorentz tensor
as long as it is a difference between connections.\ By eliminating $\mathbf{%
\omega }^{cd}$ from the Lagrangian, we are depriving $\mathbf{K}_{\ e}^{c}$
of its tensorial character; $\mathbf{K}_{\ e}^{c}$ becomes a connection, $%
\mathbf{K}_{\ e}^{c}[\mathbf{e]}=-\overset{L}{\mathbf{\omega }}{}_{\ e}^{c}[%
\mathbf{e]}$, which only keeps a tensorial behavior under \textit{global}
Lorentz transformations of the tetrad ($d\Lambda _{\ j^{\prime }}^{k}=0$ in (%
\ref{transformation})). Actually this is not a serious problem in (\ref%
{LTEGR}) because a local Lorentz transformation of the tetrad just generates
a boundary term, as could be imagined. In fact, let us perform a local
Lorentz transformation on both sides of Eq.~(\ref{EH2}) for $\mathbf{\omega }%
^{cd}=0$; since $L_{_{EH}}$ is not sensitive to a local Lorentz
transformation, then one obtains
\begin{equation}
\delta _{\Lambda }L_{_{TEGR}}=\frac{1}{4\,\kappa }\ \delta _{\Lambda
}d\left( \epsilon _{abcd}\ \mathbf{e}^{a}\ \mathbf{e}^{b}\ \mathbf{K}^{cd}[%
\mathbf{e]}\right) ,  \label{variation}
\end{equation}%
i.e.,%
\begin{equation}
\delta _{\Lambda }L_{_{TEGR}}=\frac{1}{4\,\kappa }\ d\left( \epsilon
_{abcd}\ \mathbf{e}^{a}\ \mathbf{e}^{b}\ \eta ^{de}\Lambda _{\ \ e^{\prime
}}^{c}\ d\Lambda _{\ e}^{e^{\prime }}\right) .  \label{variation2}
\end{equation}%
Therefore TEGR dynamics does not care about the local orientation of the
tetrad, meaning that TEGR, just like GR, is only involved with the dynamics of
the locally invariant metric tensor (\ref{metric}). Moreover, a boundary
term could be added to the action for balancing the behavior of $L_{_{TEGR}}$
in (\ref{variation}). In fact, we can build the strictly local Lorentz
invariant action
\begin{eqnarray}
S_{_{TEGR}}[\mathbf{e}] &=&-\frac{1}{4\,\kappa }\int\limits_{U}\epsilon
_{abcd}\ \mathbf{e}^{a}\ \mathbf{e}^{b}\ \mathbf{K}_{\ e}^{c}[\mathbf{e]\ K}%
^{ed}[\mathbf{e]}  \notag \\
&-&\frac{1}{4\,\kappa }\int\limits_{\partial U}\epsilon _{abcd}\ \mathbf{e}%
^{a}\ \mathbf{e}^{b}\ \mathbf{K}{}^{cd}[\mathbf{e],}
\label{TEGRcomplete_action}
\end{eqnarray}%
where $\mathbf{K}_{\ e}^{c}[\mathbf{e]}=-\overset{L}{\mathbf{\omega }}%
\!_{\;e}^{c}[\mathbf{e]}$. The Lagrangian (\ref{LTEGR}) is usually written as%
\begin{equation}
L_{_{TEGR}}\ =\ (2\,\kappa )^{-1}~T~\mathbf{\Omega ~,}
\end{equation}%
where $\mathbf{\Omega }$ is the metric volume $\mathbf{e}^{0}\ \mathbf{e}%
^{1}\ \mathbf{e}^{2}\ \mathbf{e}^{3}=\det [e_{\mu
}^{a}]~dx^{0}dx^{1}dx^{2}dx^{3}$, and
\begin{equation}
T=K_{\ ec}^{c}\ K_{\ \ d}^{ed}\ -K_{ed}^{c}\ K_{\ \ c}^{ed}
\end{equation}%
is the so-called Weitzenb\"{o}ck scalar. In principle, $%
T$ remains invariant only under \textit{global} Lorentz
transformations of the tetrad, since $\mathbf{K}_{\ e}^{c}$ has been
deprived of its tensor character. Expression (\ref{variation2}) was
obtained also in \cite{Rubilar} by independent means in the context
of metric affine gravity. For more details about TEGR written in the
usual index notation, see the Appendix \ref{slang}.

\section{Lorentz invariance of $f(T)$ theories}

\label{fdet}

An $f(T)$ theory consists in a deformation of the TEGR Lagrangian, as much as
an $f(R)$ theory is a deformation of the Einstein-Hilbert Lagrangian. The
teleparallel Lagrangian density $\mathcal{L}_{_{TEGR}}\ =(2\,\kappa
)^{-1}e~T $ is deformed to $\mathcal{L}=(2\,\kappa )^{-1}e~f(T)$. The
dynamical equations for $f(T)$ theories are
\begin{eqnarray}
4~e^{-1}\partial _{\mu }[e~f^{\prime }(T)~S_{a}^{\,\,\mu \nu }]
&+&4~f^{\prime }(T)~e_{a}^{\lambda }~T_{\,\,\mu \lambda }^{\rho }~S_{\rho
}^{\,\,\mu \nu }  \notag \\
-f(T)~e_{a}^{\nu } &=&-2\,\kappa \,e_{a}^{\lambda }\,\mathcal{T}_{\lambda
}^{\nu }\ ,  \label{mov}
\end{eqnarray}%
where $\mathcal{T}_{\lambda }^{\nu }$ is the energy-momentum tensor
(matter is assumed to couple the metric as usual), and
$S_{a}^{\,\,\mu \nu }$ is a quantity linear in the torsion that is
defined in the Appendix. The great advantage of field equations
(\ref{mov}) with respect the ones coming from $f(R)$ gravity,
is that they are of second order in derivatives of the dynamical field $%
\mathbf{e}^{a}$.

As an essential feature of $f(T)$ theories, the variation (\ref{variation2})
--which is essentially the variation of $T$, since the volume does not vary
under Lorentz transformations-- is trapped in the argument of function $f$
instead of being a boundary term to rule out. This feature means that the
action is sensitive to local Lorentz transformations, which implies that $%
f(T)$ theories contain dynamics not only for the metric but also for some other
degrees of freedom related to the orientation of the tetrad.\footnote{%
The presence of new degrees of freedom is also a feature characteristic of $%
f(R)$ theories (see, for instance, Ref. \cite{Sotiriou}).} Actually, because
of Eq.~(\ref{variation2}), $f(T)$ theories are invariant only under Lorentz
transformations of the tetrad accomplishing%
\begin{equation}
d(\epsilon _{abcd}\ \mathbf{e}^{a}\wedge \mathbf{e}^{b}\wedge \eta
^{de}\Lambda _{\ \ f^{\prime }}^{c}\ d\Lambda _{\ e}^{f^{\prime }})=0~.
\label{invariance}
\end{equation}%
Of course, \textit{global} Lorentz transformations ($d\Lambda _{\
e}^{f^{\prime }}=0$) do fulfill the Eq.~(\ref{invariance}). We wonder
whether the Eq.~(\ref{invariance}) has some room for a subset of \textit{%
local} Lorentz transformations. This issue is essential for understanding the
nature of the new degrees of freedom added in an $f(T)$ theory \cite{Miao}.

We shall denote $\mathcal{A}(\mathbf{e}^{a})$ the set of those local Lorentz
transformations which fulfill the equation (\ref{invariance}) for a given
frame $\mathbf{e}^{a}$, i.e, for a given solution of the field equations (%
\ref{mov}). $\mathcal{A}(\mathbf{e}^{a})$ is thus, the set of local Lorentz
transformations admissible by a certain spacetime $\mathbf{e}^{a}$, so it is
defined on shell. By virtue of the nonlinear character of (\ref{invariance}%
), it is clear that the set $\mathcal{A}(\mathbf{e}^{a})$ does
not form a Lie group in general; in fact, if $\Lambda $ and $\Lambda
^{\prime }$ belong to $\mathcal{A}(\mathbf{e}^{a})$, then the product
$\Lambda \,\Lambda ^{\prime }$ does not necessarily belong to
$\mathcal{A}(\mathbf{e}^{a})$. Nevertheless, if we have an element
of $\mathcal{A}(\mathbf{e}^{a})$ then the inverse transformation is
also in $\mathcal{A}(\mathbf{e}^{a})$; actually, since $\Lambda _{\
\ f^{\prime }}^{c}\,\Lambda _{\ e}^{f^{\prime }}=\delta _{e}^{c}$ we
can then replace $\Lambda _{\ \ f^{\prime }}^{c}\ d\Lambda _{\
e}^{f^{\prime }}$ for $-\Lambda _{\ e}^{f^{\prime }}\,d\Lambda _{\ \
f^{\prime }}^{c}$ in Eq.~(\ref{invariance}).

Let us investigate now under what circumstances the set $\mathcal{A}(\mathbf{%
e}^{a})$ becomes a Lie group. In order to do so, we shall write Lorentz
transformations as
\begin{equation}
\Lambda _{\ \ b^{\prime }}^{a}=\exp \left[ \frac{1}{2}\ \sigma ^{gh}(x)\
(M_{gh})_{\ \ b^{\prime }}^{a}\right] ,  \label{matlorentz}
\end{equation}%
where $\sigma ^{cd}(x)$ are the parameters of the transformation, and $M_{cd}
$ are six matrices labeled by antisymmetric indices that generate the
Lorentz group. The $M_{cd}$'s satisfy the algebra
\begin{equation}
\lbrack M_{ab},\,M_{cd}]=\eta _{bc}\,M_{ad}-\eta _{ac}\,M_{bd}-\eta
_{bd}\,M_{ac}+\eta _{ad}\,M_{bc}.  \label{algebra}
\end{equation}%
The components of matrices $M_{cd}$ are%
\begin{equation}
(M_{cd})_{\ \ b^{\prime }}^{a}=\delta _{c}^{a}\ \eta _{db^{\prime }}-\delta
_{d}^{a}\ \eta _{cb^{\prime }}\ .
\end{equation}%
In terms of the boost generators $K_{\alpha }=M_{0\alpha }$ and rotation
generators $J_{\alpha }=-\frac{1}{2}\epsilon _{\alpha \beta \gamma
}\,M^{\beta \gamma }$, the algebra (\ref{algebra}) is%
\begin{eqnarray}
\lbrack J_{\alpha },J_{\beta }] &=&\epsilon _{\alpha \beta \gamma }\
J^{\gamma }  \label{algebrakj} \\
\lbrack K_{\alpha },K_{\beta }] &=&-\epsilon _{\alpha \beta \gamma }\
J^{\gamma }  \notag \\
\lbrack K_{\alpha },J_{\beta }] &=&\epsilon _{\alpha \beta \gamma }\
K^{\gamma }.  \notag
\end{eqnarray}

For infinitesimal Lorentz transformations, the expression (\ref{matlorentz})
takes the form
\begin{equation}
\Lambda _{\ \ b^{\prime }}^{a}=\delta _{\ \ b^{\prime }}^{a}+\frac{1}{2}\
\sigma ^{gh}(x)\ (M_{gh})_{\ \ b^{\prime }}^{a}\ +\mathit{O}(\sigma ^{2})\ .
\end{equation}%
In this case we obtain
\begin{eqnarray}
\Lambda _{\ \ f^{\prime }}^{c}\,d\Lambda _{\ \ e}^{f^{\prime }} &\simeq &-%
\frac{1}{2}\ d\sigma ^{gh}\ (M_{gh})_{\ \ e}^{c}  \label{infinitesimal} \\
&=&-\frac{1}{2}\ d\sigma ^{gh}\ (\delta _{g}^{c}\ \eta _{he}-\delta
_{h}^{c}\ \eta _{ge})=\eta _{ge}\ d\sigma ^{gc},  \notag
\end{eqnarray}%
where we have used $\sigma ^{gh}=-\sigma ^{hg}$. Therefore, Eq.~(\ref%
{invariance}) becomes
\begin{equation}
d(\epsilon _{abcd}\ \mathbf{e}^{a}\wedge \mathbf{e}^{b}\wedge d\sigma
^{cd})\,=\,0\ ,
\end{equation}%
or, equivalently,
\begin{equation}
\epsilon _{abcd}\ d(\mathbf{e}^{a}\wedge \mathbf{e}^{b})\wedge d\sigma
^{cd}\,=\,0\ .  \label{condition}
\end{equation}%
As expected, expression (\ref{condition}) is linear in $\sigma ^{cd}$ which
means that the composition of two local infinitesimal transformations
belonging to $\mathcal{A}(\mathbf{e}^{a})$ satisfies Eq.~(\ref{condition}%
) at the lowest order in the differential of their parameters.

\bigskip

We found it very convenient to classify the solutions of the motion
equations (\ref{mov}) according to the number of closed two-forms
they involve. In this manner, a given solution $\mathbf{e}^{a}$ of
Eq.~(\ref{mov}) will be called an \emph{n-closed-area frame}
($n$-CAF), if it satisfies $d(\mathbf{e}^{a}\wedge
\mathbf{e}^{b})=0$ for $n$ of the 6 different pairs $(a$-$b)$
($0\leq n\leq6$). Clearly, from Eq.~(\ref{condition}), we have that
if $\mathbf{e}^{a}$ is a 6-CAF, then all the infinitesimal
parameters $\sigma^{cd}$ remain free. This important result just
states that for a 6-CAF, we have
$SO(3,1)_{inf}\subset\mathcal{A}(\mathbf{e}^{a})$, where
$SO(3,1)_{inf}$ stands for the infinitesimal Lorentz subgroup.

Regarding \textit{finite} transformations, from Eq.~(\ref{invariance}) it
can be proved that if two commuting local Lorentz transformations belong to $%
\mathcal{A}(\mathbf{e}^{a})$ then their composition is also an
element of $\mathcal{A}(\mathbf{e}^{a})$. Therefore, from the set
$\mathcal{A}(\mathbf{e}^{a})$ of those local Lorentz transformations
solving Eq.~(\ref{invariance}), we can extract Abelian subgroups
of the Lorentz group. Notoriously, the result (\ref{infinitesimal}),
which says that $\Lambda _{\ \ f^{\prime }}^{c}\,d\Lambda _{\ \
e}^{f^{\prime }}$ is exact at the infinitesimal level, is also
valid for separate finite boosts and rotations. As a matter of fact,
finite boosts in a given direction and rotations in a given plane
are one-parameter Lorentz transformations of the form $\Lambda =\exp
[\sigma \,M]$, where $M$ is $K_{\alpha }$ or $J_{\alpha }$ depending
on the case; therefore it is $\Lambda ^{-1}\ d\Lambda =M\,d\sigma $.
For instance, we have
\begin{equation}
\Lambda _{K_{^{3}}}^{-1}\ d\Lambda _{K_{^{3}}}~=~\left(
\begin{array}{cccc}
0 & 0 & 0 & 1 \\
0 & 0 & 0 & 0 \\
0 & 0 & 0 & 0 \\
1 & 0 & 0 & 0%
\end{array}%
\right) ~d\sigma ~,
\end{equation}%
\begin{equation}
\Lambda _{J_{^{3}}}^{-1}\ d\Lambda _{J_{^{3}}}~=~\left(
\begin{array}{cccc}
0 & 0 & 0 & 0 \\
0 & 0 & -1 & 0 \\
0 & 1 & 0 & 0 \\
0 & 0 & 0 & 0%
\end{array}%
\right) ~d\sigma ~.  \label{rotation}
\end{equation}%
Thus, Eq.~(\ref{condition}), which was obtained in the context
of infinitesimal transformations, remains valid also for separated
\emph{finite} boosts and rotations. In particular, if
$\mathbf{e}^{a}$ is a 6-CAF, then Eq.~(\ref{invariance}) will be
satisfied for any local boost or rotation. This remark seems to
indicate that the finite local transformations will be organized in
6 Abelian subgroups of dimension 1 (each corresponding to a
boost in a given direction or a rotation in a given plane). However
we will show below that for a given n-CAF, a number
$\lfloor\frac{n}{2}\rfloor$ of two-dimensional Abelian subgroups of
the type $\{K_{\alpha },\,J_{\alpha }\}$ can also appear (here
$\lfloor\,\rfloor$ refers to the floor function). For $n\geq4$ their
appearance will actually be unavoidable.

\bigskip

In order to proceed constructively, let us begin by considering a
1-CAF such that, let us say, $d(\mathbf{e}^{0}\wedge
\mathbf{e}^{3})=0$. This property implies that the local parameter
$\sigma ^{12}$ can be freely chosen without affecting the
fulfillment of Eq.~(\ref{condition}). As said, this result is also
valid for \textit{finite} local rotations generated by
$M_{12}=-J_{3}$. In fact, Eq.~(\ref{rotation}) shows that the
exact matrix-valued 1-form $\Lambda
_{(J^{3})}^{-1}\ d\Lambda _{(J^{3})}$ only contributes to Eq.~(\ref%
{invariance}) through the components $(1$-$2)$; however such a
contribution is canceled whenever $d(\mathbf{e}^{0}\wedge
\mathbf{e}^{3})$ vanishes. We then get a one-dimensional subgroup of
finite local transformations (the subgroup of rotations about
$x^3$). This reasoning is applicable to any of the other possible
closed areas as well.

In general, for an $n$-CAF one could expect $n$ one-dimensional subgroups
of finite local transformations. However, if $n\geq 2$ there is a more
interesting case. Let us consider the case $d(\mathbf{e}^{0}\wedge \mathbf{e}%
^{3})=0=d(\mathbf{e}^{1}\wedge \mathbf{e}^{2})$. Then Eq.~(\ref{condition})
is accomplished by local transformations generated by combinations of $M_{12}
$ and $M_{03}$ (i.e., $J_{3}$ and $K_{3}$). Since these \textit{commuting}
generators preserve the closedness of \textit{both} areas, we can expect
that the result remains valid for finite local transformations generated by $%
M_{12}$ and $M_{03}$.\ In fact, if $\Lambda $ is
\begin{equation}
\Lambda ~=~\left(
\begin{array}{cccc}
\cosh \sigma  & 0 & 0 & \sinh \sigma  \\
0 & \cos \alpha  & -\sin \alpha  & 0 \\
0 & \sin \alpha  & \cos \alpha  & 0 \\
\sinh \sigma  & 0 & 0 & \cosh \sigma
\end{array}%
\right) ~,
\end{equation}%
then it will be%
\begin{equation}
\Lambda ^{-1}\ d\Lambda ~=~\left(
\begin{array}{cccc}
0 & 0 & 0 & d\sigma  \\
0 & 0 & -d\alpha  & 0 \\
0 & d\alpha  & 0 & 0 \\
d\sigma  & 0 & 0 & 0%
\end{array}%
\right) ~.
\end{equation}%
So two independent local parameters $\sigma(x^{\mu}) $ and $\alpha( x^{\mu})$ can be chosen
without affecting the fulfillment of Eq.~(\ref{invariance}), because they
contribute just to terms that are canceled by the vanishing of $d(\mathbf{e}%
^{0}\wedge \mathbf{e}^{3})$ and $d(\mathbf{e}^{1}\wedge
\mathbf{e}^{2})$. So we get a two-dimensional Abelian subgroup (we
have $\lfloor\frac{n}{2}\rfloor=1$ in this case). Schematically, we
have then
\begin{eqnarray}\label{abgroups}
d(\mathbf{e}^{0}\wedge \mathbf{e}%
^{1})=0=d(\mathbf{e}^{2}\wedge \mathbf{e}^{3})&\rightarrow &\{K_{1},\,J_{1}\}\notag\\
d(\mathbf{e}^{0}\wedge \mathbf{e}%
^{2})=0=d(\mathbf{e}^{1}\wedge \mathbf{e}^{3})&\rightarrow &\{K_{2},\,J_{2}\}\notag\\
d(\mathbf{e}^{0}\wedge \mathbf{e}%
^{3})=0=d(\mathbf{e}^{1}\wedge \mathbf{e}^{2})&\rightarrow &\{K_{3},\,J_{3}\}.
\end{eqnarray}%
However, there exist other types of 2-CAFs, for instance, the one
having $d(\mathbf{e}^{0}\wedge
\mathbf{e}^{1})=0=d(\mathbf{e}^{0}\wedge \mathbf{e}^{2})$. This
2-CAF will lead to free $M_{01}=K_{1}$ and $M_{02}=K_{2}$, but these
do not commute. So, the appearance or not of a two-dimensional Abelian
subgroup of the Lorentz group in a 2-CAF, depends on the closed
areas it involves. It can be checked that this is also true for a
3-CAF. In this case, if the 3-CAF involves the proper closed areas,
we also expect just one two-dimensional Abelian subgroup
($\lfloor\frac{3}{2}\rfloor=1$).

If $n\geq 4$ the emergence of two-dimensional Abelian subgroups is unavoidable. In the case $n=4,5$ we shall obtain two of
them, and for $n=6$ we will obtain the maximum number of such groups, i.e., three; these will be just $\{K_{1},\,J_{1}\}$,
$\{K_{2},\,J_{2}\}$, and $\{K_{3},\,J_{3}\}$. It should be noticed
that these subgroups cannot be combined in a larger
group: only one of them can be locally applied to the solution $\mathbf{e}%
^{a}$ while the rest of the symmetries remain global. This is so because the
local action of one of them will affect the closedness of the rest of the
closed areas.

\bigskip

For $n\geq 4$, let us consider case including the subgroups $%
\{K_{1},\,J_{1}\}$, $\{K_{2},\,J_{2}\}$ (i.e., $d(\mathbf{e}^{0}\wedge
\mathbf{e}^{1})=0=d(\mathbf{e}^{2}\wedge \mathbf{e}^{3})$ and $d(\mathbf{e}%
^{0}\wedge \mathbf{e}^{2})=0=d(\mathbf{e}^{1}\wedge \mathbf{e}^{3})$). In
such case there is another way of organizing the subgroups. In fact, we can
introduce the Abelian subgroups $\{A^{(1)},\,A^{(2)}\}$, $%
\{B^{(1)},\,B^{(2)}\}$, where
\begin{eqnarray}
A^{(1)} &\doteq &K_{1}+J_{2}\ ,\,\,\,\,\,\,\,A^{(2)}\doteq K_{2}-J_{1}\ ,
\notag \\
B^{(1)} &\doteq &K_{1}-J_{2}\ ,\,\,\,\,\,\,\,B^{(2)}\doteq K_{2}+J_{1}\ .
\end{eqnarray}%
The Lorentz transformation generated by $A^{(1)}$ is $\Lambda
_{A^{(1)}}=\exp [\sigma \ A^{(1)}].$ So the matrix $\Lambda _{\ \ f^{\prime
}}^{c}\ d\Lambda _{\ e}^{f^{\prime }}$ in Eq.~(\ref{invariance}) is%
\begin{equation}
\Lambda _{A^{(1)}}^{-1}\ d\Lambda _{A^{(1)}}~=~A^{(1)}~d\sigma ~~=~\left(
\begin{array}{cccc}
0 & 1 & 0 & 0 \\
1 & 0 & 0 & 1 \\
0 & 0 & 0 & 0 \\
0 & -1 & 0 & 0%
\end{array}%
\right) ~d\sigma ~.
\end{equation}%
The rest of the cases are obtained by using the matrices%
\begin{eqnarray}
A^{(2)}&=&~\left(
\begin{array}{cccc}
0 & 0 & 1 & 0 \\
0 & 0 & 0 & 0 \\
1 & 0 & 0 & 1 \\
0 & 0 & -1 & 0%
\end{array}%
\right) , \notag\\
B^{(1)}&=&~\left(
\begin{array}{cccc}
0 & 1 & 0 & 0 \\
1 & 0 & 0 & -1 \\
0 & 0 & 0 & 0 \\
0 & 1 & 0 & 0%
\end{array}%
\right),\notag\\
B^{(2)}&=&~\left(
\begin{array}{cccc}
0 & 0 & 1 & 0 \\
0 & 0 & 0 & 0 \\
1 & 0 & 0 & -1 \\
0 & 0 & 1 & 0%
\end{array}%
\right) .
\end{eqnarray}
As can be seen, the contributions of any of these local transformations to Eq.~(\ref{invariance}) will be canceled by the closedness of the areas $%
(0$-$1)$, $(2$-$3)$, $(0$-$2)$, and $(1$-$3)$. It is worth noticing that $%
\{A^{(1)},\,A^{(2)}\}$ ($\{B^{(1)},\,B^{(2)}\}$) constitute the
Abelian sector of the little group for massless particles traveling
towards decreasing (increasing) values of $x^{3}$ \cite{Wigner}.
\footnote{The little group also includes the rotations generated by
$J_{3}$. Thus, its algebra gets the form of the algebra of
translations and rotations in the Euclidean plane. It has been
proved that $A^{(1,2)}$, $B^{(1,2)}$ generate gauge transformations
of the electromagnetic field \cite{Janner,Kim}.}

\bigskip

We conclude this Section with two remarks. Of course, the classification of
tetrads through the number of closed areas they contain is not invariant
under global Lorentz transformations. Actually we can use the always admissible global Lorentz
transformations to maximize the number $n$ for the tetrad under
consideration. Besides, the scheme of $n-$CAFs does not exhaust the
chances of obtaining a local invariance for a given solution of the
equations of motion (\ref{mov}). For instance, even if $\mathbf{e}^{0}\wedge
\mathbf{e}^{1}$ were not closed, $\sigma ^{23}$ could admit a limited
dependence on the coordinates without destroying the validity of Eq.~(\ref%
{condition}). This means, on one hand, that even for a 0-CAF the
possibility of a restricted local invariance is still present, and
on the other, that the remnant group for a given $n-$CAF can be
larger than that considered in the paragraphs above. This restricted
local invariance depends on the form of each solution and it should
be considered in each particular case, as we will show in the next
Section.

\bigskip

\section{Examples}

\label{sec:infini}

In this section we will offer a number of simple but quite important
examples that will help to visualize the ideas displayed in the preceding
paragraphs.

\subsection{Minkowski spacetime}

\label{minkowski}

Perhaps one of the most important cases to be analyzed should be
Minkowski spacetime, because it approximately represents the
geometrical arena where our daily experience takes place. For this
reason it is our concern now to figure out what kind of local
Lorentz transformations we are free to perform in the Euclidean
frame (see below), in order to be unable to distinguishing them
from the outcomes of experiments performed in our local lab.

The Euclidean frame $\mathbf{e}^{a}=\delta _{\,\,b}^{a}\,dx^{b}$ is a global
smooth basis for Minkowski spacetime (the $x^{b}$'s refer to $x^{0,\alpha}$,
where $x^{\alpha}$ are Cartesian coordinates). Since $T^{a}=d\mathbf{e}%
^{a}=0 $, the Weitzenb\"{o}ck scalar is identically null, and the Euclidean
frame is a vacuum solution of equation (\ref{mov}) for any $f(T)$ function
smooth at $T=0$ \cite{Nos4}.\footnote{%
Other $f(T)$ deformations of GR, such as the ones used for describing the late
time cosmic speed-up (for instance $f(T)=T+\alpha /T$), do not have Minkowski
spacetime as a vacuum solution. Instead, they lead to a constant but non-
null $T$, and so, to a de Sitter or anti de Sitter spacetime.}

The Euclidean frame is perhaps the best example of a 6-CAF. Therefore, $f(T)$ theories that
are smooth at $T=0$ do not distinguish among locally related orthonormal
frames in Minkowski spacetime. In other words, the absence of gravity in
$f(T)$ theories is revealed as an incapability in the selection of a
preferred parallelization at a local level.

\bigskip

\subsection{Cosmological spacetimes}

\subsubsection{Spatially flat Friedmann-Robertson-Walker spacetimes}

\label{frwplano}

The diagonal frame $\mathbf{e}^{0}=dt$, $\mathbf{e}^{\alpha
}=a(t)~\delta _{i}^{\alpha }dx^{i}$ is a solution to Eq.~(\ref{mov})
for flat Friedmann-Robertson-Walker (FRW) spacetimes \cite{Nos3}.
This frame is a 3-CAF since $d(\mathbf{e}^{0}\wedge
\mathbf{e}^{\alpha })=0$, $\forall \alpha $. Because
of the comments made in the last section, we expect $\mathcal{A}(%
\mathbf{e}^{a})$ to include at least three one-dimensional Abelian subgroups of the Lorentz group.
Actually, equation (\ref{invariance}) is accomplished for
any local rotation $\sigma ^{\beta \gamma }(x^{a})$ of the diagonal frame $%
\mathbf{e}^{a}$, because for every pair $(0\,\alpha )$, we have a pair $%
(\beta \,\gamma )$ (since these last two are different from $\alpha $), and
there are three such pairs. Then, $\mathcal{A}(%
\mathbf{e}^{a})$ includes the three Abelian subgroups of rotation about a given axis.

A nice example of the behavior discussed at the end of the last section
(i.e., the presence of an admissible transformation even though $\mathbf{e}%
^{\alpha }\wedge \mathbf{e}^{\beta }$ is not closed), can be shown as
follows. Since we have
\begin{equation}
d(\mathbf{e}^{\alpha }\wedge \mathbf{e}^{\beta })=2a\,\overset{\cdot }{a}%
\,dt\wedge dx^{\alpha }\wedge dx^{\beta },  \label{cosmo3}
\end{equation}%
we note that three Lorentz boosts $\sigma ^{0\gamma }(t,x^{\alpha },x^{\beta
})$ of $\mathbf{e}^{a}$ will also lead to an equivalent solution of the
dynamical equations (\ref{mov}) (take note that $\gamma \neq \alpha \neq
\beta $). This is so because the 1-form $d\sigma ^{0\gamma }$ in Eq.~(\ref%
{condition}) does not contain a term proportional to $dx^{\gamma }$, so the
wedge product $d(\mathbf{e}^{\alpha }\wedge \mathbf{e}^{\beta })\wedge
d\sigma ^{0\gamma }$ is null. Then, for this particular 3-CAF, we have that $%
\mathcal{A}(\mathbf{e}^{a})$ contains not three, but six independent
generators $\sigma ^{0\gamma }(t,x^{\alpha },x^{\beta })$ and $%
\sigma ^{\beta \gamma }(x^{\mu})$. Nevertheless, it is important to
realize that the three one-dimensional Abelian subgroups of boosts in
a given direction (generated by $\sigma ^{0\gamma }(t,x^{\alpha
},x^{\beta })$), are constrained to possess a restricted dependence
on the spacetime coordinates $x^{\mu}$. For instance, if we consider
a boost in the $t-x$ plane, we have that the generator can depend
just on $(t,y,z)$. For boosts in the remaining planes, analogous
comments are in order.

Because of this, even though this particular 3-CAF does not allow
the emergence of a two-dimensional Abelian subgroup of the form
$\{K_{\alpha}(x^{\mu}),J_{\alpha}(x^{\mu})\}$ (as explained in the
paragraph below Eq.~(\ref{abgroups})), we still expect three Abelian
subgroups of dimension 2 with restricted coordinate dependence
contained in $\mathcal{A}(\mathbf{e}^a)$. Precisely, these are the
ones generated by
\begin{eqnarray}\label{restricted}
&&\{K_{x}(t,y,z),J_{x}(x^{\mu})\}  \notag
\\
&&\{K_{y}(t,x,z),J_{y}(x^{\mu})\}  \notag
\\
&&\{K_{z}(t,x,y),J_{z}(x^{\mu})\}.
\end{eqnarray}
Spatially flat FRW cosmological models admit, hence, an infinite
number or proper tetrads, organized in the Abelian subgroups of the
Lorentz group just mentioned. This strong result seems to suggest that
some claims present in the literature regarding superluminal
propagating modes and nonuniqueness of time evolution in $f(T)$
theories \cite{nester1,nester2} should be revised in light of the new developments here introduced.

\subsubsection{Spatially curved FRW spacetime}

The parallelization of closed and open FRW universes is much less obvious.
In these cases we can write the line element as
\begin{equation}
ds^{2}=dt^{2}-a^{2}(t)k^{2}[d(k\psi )^{2}+\sin ^{2}(k\psi )(d\theta
^{2}+\sin ^{2}\theta \,d\phi ^{2})],  \label{metcurv}
\end{equation}%
where $(\psi ,\theta ,\phi )$ are standard hyperspherical coordinates on
the three-sphere. The parameter $k$ appearing in (\ref{metcurv}) takes the
values $k=1$ for the spatially spherical universe and $k=i$ for the
spatially hyperbolic one.

In Ref.~\cite{Nos3} it was shown how one can find a global frame for
spatially curved FRW spacetimes, i.e., a global basis that turns the
dynamical equations (\ref{mov}) into a consistent system of
differential equations for the scale factor $a(t)$. It reads
\begin{equation}
 \mathbf{e}^{0}=dt,\,\,\,\,\,\mathbf{e}^{\alpha }=a(t)~%
\mathbf{E}^{\alpha}, \label{frameK=1}
\end{equation}%
where the 1-forms $\mathbf{E}^{\alpha }$ are
\begin{eqnarray}  \label{framecurvo}
\frac{\mathbf{E}^{1}}{k} &=&-k\cos \theta \,d\psi +\sin \theta \,\sin (k\psi
)\cos (k\psi )\,d\theta - \\
&&-\sin ^{2}(k\psi )\sin ^{2}\theta \,d\phi ,  \notag \\
\frac{\mathbf{E}^{2}}{k} &=&\,\,\,\,k\sin \theta \cos \phi \,d\psi -  \notag
\\
&&-\sin ^{2}(k\psi )[\sin \phi -\cot (k\psi )\cos \theta \cos \phi
]\,d\theta -  \notag \\
&&-\sin ^{2}(k\psi )\sin \theta \lbrack \cot (k\psi )\sin \phi +\cos \theta
\cos \phi ]\,d\phi ,  \notag \\
\frac{\mathbf{E}^{3}}{k} &=&-k\sin \theta \sin \phi \,d\psi -  \notag \\
&&-\sin ^{2}(k\psi )[\cos \phi +\cot (k\psi )\cos \theta \sin \phi
]\,d\theta -  \notag \\
&&-\sin ^{2}(k\psi )\sin \theta \lbrack \cot (k\psi )\cos \phi -\cos \theta
\cos \phi ]\,d\phi .  \notag
\end{eqnarray}
Many fewer local symmetries are left in this case, because the frame
(\ref{frameK=1}) is just a 0-CAF. Nonetheless, since
$d(\mathbf{e}^{0}\wedge \mathbf{e}^{\alpha })=dt\wedge
d\mathbf{e}^{\alpha }$, we can say that time-dependent rotations
$\sigma ^{\beta \gamma }(t)$ are authorized by equation
(\ref{condition}). Thus, we get three one-dimensional Abelian subgroups
composed of time-dependent rotations about a given axis.

\subsubsection{Bianchi type I models}

Homogeneous and anisotropic Bianchi type I models are described by the line
element
\begin{equation}
ds^{2}=dt^{2}-a_{1}^{2}(t)\,dx^{2}-a_{2}^{2}(t)\,dy^{2}-a_{3}^{2}(t)\,dz^{2}.
\label{metcurv}
\end{equation}%
The manifold topology is $R^{4}$, so a proper parallelization is given by
the frame $\mathbf{e}^{0}=dt$, $\mathbf{e}^{1}=a_{1}\,dx$, $\mathbf{e}%
^{2}=a_{2}\,dy$, $\mathbf{e}^{3}=a_{3}\,dz$. Although Bianchi type I
spacetimes contain less isometries than FRW cosmologies, we can easily check
that $d(\mathbf{e}^{0}\wedge \mathbf{e}^{\alpha })=0$, $\forall \alpha $, so
we are in the presence of a 3-CAF once again, and the same comments of
section \ref{frwplano} are in order.

\section{Concluding comments}

\label{sec:conclusion}

For special kind of frames, the so-called $n$-CAFs (which include
Minkowski spacetime and a wide variety of cosmological models), we have
obtained in section \ref{fdet} a number of results regarding certain
conditions for a local Lorentz transformation that belongs to the set $\mathcal{A}%
(\mathbf{e}^a)$. In particular, for 6-CAFs, we have concluded the following:

\begin{description}
\item[1.] \emph{Any} infinitesimal local Lorentz transformation belongs to $%
\mathcal{A}(\mathbf{e}^{a})$. \label{item1}

\item[2.] Regarding finite transformations, we have that $\langle
K_{\alpha}(x^{\mu}),J_{\alpha}(x^{\mu})\rangle\subset\mathcal{A}(\mathbf{e}^a)$,
where $K_{\alpha}(x^{\mu}),J_{\alpha}(x^{\mu})$ are the six
generators of the two-dimensional Abelian subgroups of the Lorentz
group. In particular, six one-dimensional Abelian subgroups are
included in $\mathcal{A}(\mathbf{e}^a)$ (boosts in a given direction
and rotations about a given axis).  \label{item2}
\end{description}

As a direct consequence, we see that there are infinitely many
adequate tetrads representing Minkowski spacetime in $f(T)$ gravity.
This result does not mean that any tetrad giving rise to the
Minkowski metric is a solution of the $f(T)$ motion equations in
vacuum. For instance, the tetrad
\begin{equation}
\mathbf{e}^0=dt,\,\,\,\,\mathbf{e}^1=dr,\,\,\,\,\mathbf{e}^2=r\,
d\theta,\,\,\,\,\mathbf{e}^3=r\, \sin\theta \,d\phi  \label{tetradain}
\end{equation}%
corresponding to $ds^2=dt^2-dr^2-r^2(d\theta^2+\sin^2\theta\, d\phi^2)$, is
not a solution of the vacuum field equations, because it fails to be a basis
of the tangent space at $r=0$, and so, it is not a parallelization of
Minkowski spacetime. This is so because (\ref{tetradain}) can be obtained
from the Euclidean frame by means of a local Lorentz transformation which is
not an element of $\mathcal{A}(\mathbf{e}^a)$. Precisely, the Euclidean
frame in spherical coordinates stands as (just change coordinates in $\mathbf{e}%
^{a}=\delta _{\,\,b}^{a}\,dx^{b}$)
\begin{eqnarray}
\mathbf{e}^{0} &=&dt,  \notag \\
\mathbf{e}^{1} &=&\sin\theta \cos\phi \ dr+r\cos \theta \cos \phi \ d\theta
-r\sin \theta \sin \phi \ d\phi ,  \notag \\
\mathbf{e}^{2} &=&\sin \theta \sin \phi \ dr+r\cos \theta \sin \phi \
d\theta +r\sin \theta \cos \phi \ d\phi,  \notag \\
\mathbf{e}^{3} &=&\cos\theta \ dr-r\sin \theta \ d\theta ,  \label{ansatz}
\end{eqnarray}%
The tetrad (\ref{tetradain}) can be obtained from (\ref{ansatz}) by
means of a local rotation which, however, does not satisfy
Eq.~(\ref{invariance}) because the involved generators do not
commute.

For $n$-CAFs (flat FRW and Bianchi type I models of section \ref{sec:infini}
being in this category), the picture is more restrictive, and the statements
made for 6 CAFs change to

\begin{description}
\item[1$^{\prime }$.] \emph{Some} infinitesimal local Lorentz
transformations, generated by $n$ one-dimensional subgroups of the infinitesimal
Lorentz group, belong to $\mathcal{A}(\mathbf{e}^{a})$. \label{item1prima}

\item[2$^{\prime}$.] Regarding finite transformations, a number $\lfloor n/2\rfloor$ of two-dimensional Abelian subgroups
of the form $K_{\alpha}(x^{\mu}),J_{\alpha}(x^{\mu})$ might arise,
depending on the particular closed area involved. For $n\geq4$ these
Abelian subgroups will actually exist. In particular $n$ one-dimensional Abelian subgroups will be included always in
$\mathcal{A}(\mathbf{e}^a)$. Sometimes, depending on the specific
form of the $n$-CAF, an additional (restricted) Lorentz invariance
can exist (see, e.g., Eq.~(\ref{restricted})). \label{item2prima}
\end{description}

\bigskip

Finally, we would like to mention some remaining open questions of
conceptual guise. One of these, concerns the relationship between
the isometries of a given spacetime
$(\mathcal{T}(\mathcal{M}),\mathbf{e}^{a})$, and its remnant set
$\mathcal{A}(\mathbf{e}^a)$. Perhaps it would be plausible to think
that an increase in the number of isometries will lead to an
enlargement of the set $\mathcal{A}(\mathbf{e}^a)$. It should be
clear from the examples examined above that this is not actually
true. All FRW
spacetimes have the same number of isometries, whereas the set $\mathcal{A}(%
\mathbf{e}^a)$ is considerably larger for spatially flat models. More
drastically, curved FRW spacetimes have a notoriously smaller $\mathcal{A}(%
\mathbf{e}^a)$ compared with the less symmetric Bianchi type I models.

Presumably, the answer to this issue underlies the global properties of
the cited spacetimes and not only in their local geometry. As a matter of
fact, flat FRW and Bianchi type I spacetimes both have topology $R^4$, and
they are both represented by 3-CAFs. In turn, due to that fact that (let us
say) closed FRW spacetimes have topology $R\times S^3$, we should expect a
more involved global behavior concerning the parallelization process, which
reflects itself in the fact that the frame (\ref{framecurvo}) is just a
0-CAF.

As a final remark, we can comment on an important result obtained in
Refs. \cite{Heyde,Hartley}. There it was shown that where the
connection is other than the Levi-Civita connection, the notion of an inertial reference frame can still be defined locally by means of
local normal frames. This is a realization of the equivalence
principle in theories with torsion, which means that in a spacetime
with an arbitrary (though metric compatible) connection, we still
recover the Minkowskian behavior locally. It would be interesting to
figure out under what circumstances this property will still hold
for (torsional) theories of gravity in which the Lorentz symmetry is
not fully present, in the sense discussed in this work. By virtue of
the result here obtained, the existence of locally inertial frames
would assure (locally) the full Lorentz symmetry of any spacetime
arising as a solution of the $f(T)$ field equations, an so a well-behaved causal structure at a local level \cite{zeeman}.
\bigskip

\emph{Acknowledgements}. We would like to dedicate this work to the memory of Professor
Rafael Montemayor. This work was supported by CONICET, Universidad de Buenos Aires, and Instituto Balseiro.

\section{Appendix}
\subsection{On the contorsion tensor}\label{propcontor}

Some properties of the contorsion tensor can be enumerated as
follows:

\begin{enumerate}
  \item The equation of geodesics in an arbitrary connection is $(D%
\mathbf{U}/D\tau )^{i}=\mathbf{K}_{\ j}^{i}(\mathbf{U})\,U^{j}$, so the
contorsion represents the gravitational force.
  \item $\mathbf{T}^{i}=\mathbf{K}_{\ j}^{i}\wedge \mathbf{E}^{j}$,
then $\mathbf{(T}^{i})_{jk}=(\mathbf{K}_{\;k}^{i})_{j}-(\mathbf{K}%
_{\;j}^{i})_{k}$ (combine Eqs.~(\ref{torsion}), (\ref{Levi-Civita}) and (\ref%
{contorsion})).
  \item If $\overset{L}{D}$ is the covariant derivative associated with the
  Levi-Civita connection (\ref{Levi-Civita}), then it results\\
$D\mathbf{K}_{\;j}^{i}-\overset{L}{D}\mathbf{K}_{\;j}^{i}=2\
\mathbf{K}_{\;k}^{i}\wedge \mathbf{K}_{\;j}^{k}\ $.
  \item $\mathbf{R}^{i}_{\;j}-\overset{L}{\mathbf{R}}{}^{i}_{\ j}=%
\overset{L}{D}\mathbf{K}_{\ j}^{i}+\mathbf{K}_{\ k}^{i}\wedge \mathbf{K}_{\
j}^{k}=D\mathbf{K}_{\ j}^{i}-\mathbf{K}_{\ k}^{i}\wedge \mathbf{K}_{\ j}^{k}$.
  \item $\mathbf{K}_{ba}=-\mathbf{K}_{ab}$ (use (\ref{metricity})).
  \item $K_{abc}\doteq \left( \mathbf{K}_{ab}\right) _{c}=-\frac{1}{2}%
\ \left[ \left( \mathbf{T}_{a}\right) _{bc}+\left( \mathbf{T}_{b}\right)
_{ca}-\left( \mathbf{T}_{c}\right) _{ab}\right] =-\ \frac{1}{2}\ \left[
\left( D\mathbf{e}_{a}\right) _{bc}+\left( D\mathbf{e}_{b}\right)
_{ca}-\left( D\mathbf{e}_{c}\right) _{ab}\right] $\ (use Property 2 and Eq.~(\ref%
{torsion})).
\end{enumerate}

\subsection{TEGR in usual language}\label{slang}

The Weitzenb\"{o}ck connection is defined in a given orthonormal basis $\{\mathbf{e}%
^{a}\}$ as $\mathbf{\omega }^{cd}=0$. Since $\mathbf{e}^{a}=e_{\mu }^{a}\
dx^{\mu }$, one realizes that the transformation between coordinate and
orthonormal bases uses the coefficients $\Lambda _{\mu }^{a}=e_{\mu }^{a}$.
According to (\ref{transformation}), if the connection vanishes in the basis
$\{\mathbf{e}^{a}\}$ then it transforms to a coordinate basis as $(\mathbf{%
\omega }_{\ \nu }^{\mu })_{\lambda }=e_{a}^{\mu }\ \partial _{\lambda
}e_{\nu }^{a}$, which is the familiar form of Weitzenb\"{o}ck connection. In
Weitzenb\"{o}ck spacetime, it is $T^{\mu }=D\ dx^{\mu }=\mathbf{\omega }_{\ \nu
}^{\mu }\wedge dx^{\nu }=e_{a}^{\mu }\ \partial _{\lambda }e_{\nu }^{a}\
dx^{\lambda }\wedge dx^{\nu }$, i.e.,%
\begin{equation}
T_{\ \lambda \nu }^{\mu }=e_{a}^{\mu }\ (\partial _{\lambda }e_{\nu
}^{a}-\partial _{\nu }e_{\lambda }^{a})~.  \label{Wtorsion}
\end{equation}%
To recover the familiar form of $L_{_{TEGR}}$ one writes $\mathbf{K}_{\
e}^{c}=K_{\ ef}^{c}\ \mathbf{e}^{f}$, so%
\begin{equation}
L_{_{TEGR}}\ =\ -\frac{1}{4\,\kappa }\ K_{\ ef}^{c}\ K_{\ \ g}^{ed}\
\epsilon _{abcd}\ \mathbf{e}^{a}\ \mathbf{e}^{b}\ \mathbf{e}^{f}\ \mathbf{e}%
^{g}\ .
\end{equation}%
In this expression one recognizes the volume 4-form $\mathbf{\Omega }$,%
\begin{equation}
\mathbf{e}^{a}\ \mathbf{e}^{b}\ \mathbf{e}^{f}\ \mathbf{e}^{g}=\epsilon
^{abfg}\ \mathbf{\Omega }=\epsilon ^{abfg}\ e\ \ dx^{0}\ dx^{1}\ dx^{2}\
dx^{3}\ ,
\end{equation}%
where $e\doteq \det [e_{\mu }^{a}]$. We use the identity $\epsilon ^{abfg}\
\epsilon _{abcd}=-2(\delta _{c}^{f}\delta _{d}^{g}-\delta _{c}^{g}\delta
_{d}^{f})$ to obtain%
\begin{equation}
L_{_{TEGR}}\ =\ \frac{1}{2\,\kappa }\ (K_{\ ec}^{c}\ K_{\ \ d}^{ed}\ -K_{\
ed}^{c}\ K_{\ \ c}^{ed})\ \mathbf{\Omega }\ .
\end{equation}%
According to property 4 of appendix \ref{propcontor} it is $K_{\ ec}^{c}=-T_{\ ec}^{c}$, $K_{\ \
d}^{ed}=-T_{\ \;d}^{d\ \;e}$ (we exploited the antisymmetry of torsion).
Also $K_{\ ed}^{c}\ K_{\ \ c}^{ed}=K_{\ [ed]}^{c}\ K_{\ \ c}^{ed}=(-1/2)\
T_{\ ed}^{c}\ K_{\ \ c}^{ed}$. Then,
\begin{eqnarray*}
&&K_{\ ec}^{c}\ K_{\ \ d}^{ed}\ -K_{\ ed}^{c}\ K_{\ \ c}^{ed}=T_{\ ec}^{c}\
T_{\ \;d}^{d\;\ e}+\frac{1}{2}\ T_{\ ed}^{c}\ K_{\ \ c}^{ed} \\
&=&\frac{1}{2}\ T_{\ ed}^{c}(T_{a}^{\ ae}\delta _{c}^{d}-T_{a}^{\ ad}\delta
_{c}^{e}+K_{\ \ c}^{ed})=T_{\ ed}^{c}\ S_{c}^{\ ed}\ ,
\end{eqnarray*}%
where%
\begin{equation}
S_{c}^{\ ed}\doteq \frac{1}{2}\ K_{\ \ c}^{ed}+T_{a}^{\ a[e}\delta _{c}^{d]}=%
\frac{1}{2}\ K_{\ \ c}^{ed}+K_{a}^{\ a[e}\delta _{c}^{d]}\ .  \label{S}
\end{equation}%
The quantity $T_{\ ed}^{c}\ S_{c}^{\ ed}$ is the Weitzenb\"{o}ck scalar $T$. All
these quantities behave tensorially under local Lorentz transformations
whenever the spin connection is not frozen to zero. Otherwise, they are
tensors just under global Lorentz transformations.

\bigskip The boundary term in (\ref{TEGRcomplete_action}) contributes $%
-(4\,\kappa )^{-1}d(\epsilon _{abcd}\ \mathbf{e}^{a}\ \mathbf{e}^{b}\
\mathbf{K}{}^{cd})$ to the Lagrangian. This exact $4-$form can be rewritten
in terms of a four-divergence. Notice that%
\begin{equation*}
d(\epsilon _{abcd}\ \mathbf{e}^{a}\ \mathbf{e}^{b}\ \mathbf{K}%
{}^{cd})=d(\epsilon _{abcd}\ K_{\ \ e}^{cd}\ \mathbf{e}^{a}\ \mathbf{e}^{b}\
\mathbf{e}^{e})\ ,
\end{equation*}%
where%
\begin{equation*}
\mathbf{e}^{a}\ \mathbf{e}^{b}\ \mathbf{e}^{e}=-\epsilon ^{abef}\ \mathbf{%
\Omega (e}_{f})\ .
\end{equation*}%
Therefore%
\begin{eqnarray}
&&d(\epsilon _{abcd}\ \mathbf{e}^{a}\ \mathbf{e}^{b}\ \mathbf{K}%
{}^{cd})=d\left( 2(\delta _{c}^{e}\delta _{d}^{f}-\delta _{c}^{f}\delta
_{d}^{e})\ K_{\ \ e}^{cd}\ \mathbf{\Omega (e}_{f})\right)  \notag \\
&=&4\ d\left( K_{\ \ c}^{cd}\ \mathbf{\Omega (e}_{d})\right) =4\ div(K_{\ \
c}^{cd}\ \mathbf{e}_{d})\ \mathbf{\Omega }  \notag \\
&=&\frac{4}{e}\ \partial _{\mu }(e\ K_{\ \ c}^{cd}\ e_{d}^{\mu })\ \mathbf{%
\Omega }\ .  \label{divergence}
\end{eqnarray}%
According to (\ref{LC-components}) it is%
\begin{eqnarray*}
K_{\ \ c}^{cd}[\mathbf{e]} &=&-\left( d\mathbf{e}^{c}\right) _{~c}^{d}=\eta
^{db}~\partial _{\lambda }e_{\nu }^{c}~(e_{c}^{\lambda }~e_{b}^{\nu
}-e_{b}^{\lambda }~e_{c}^{\nu }) \\
&=&\eta ^{db}~e_{c}^{\lambda }~e_{b}^{\nu }~(\partial _{\lambda }e_{\nu
}^{c}-\partial _{\nu }e_{\lambda }^{c}).
\end{eqnarray*}%
By comparing with equation (\ref{Wtorsion}), one obtains%
\begin{equation*}
d(\epsilon _{abcd}\ \mathbf{e}^{a}\ \mathbf{e}^{b}\ \mathbf{K}{}^{cd})=\frac{%
4}{e}\ \partial _{\mu }\left( e~T_{\ \;\lambda }^{\lambda \;\ \mu }\right) \
\mathbf{\Omega }\ .
\end{equation*}

\end{document}